\documentclass[structabstract]{aa}  
\usepackage{natbib}
\bibpunct{(}{)}{;}{a}{}{,} 
\usepackage{graphicx}

\usepackage{txfonts}
\begin{document}

\newcommand{\kms}{km s$^{-1}$}
\title{Activity in galactic nuclei of cluster and field galaxies in the local universe}
\author{H. S. Hwang \inst{1,2}
\and C. Park \inst{3}
\and D. Elbaz \inst{1}
\and Y.-Y. Choi \inst{4}
}
\institute{CEA Saclay/Service d'Astrophysique, F-91191 Gif-sur-Yvette, France \\
\email{hhwang@cfa.harvard.edu;delbaz@cea.fr}
\and Smithsonian Astrophysical Observatory, 60 Garden Street, Cambridge, MA 02138, USA
\and School of Physics, Korea Institute for Advanced Study, Seoul 130-722, Korea
\email{cbp@kias.re.kr} 
\and Dept. of Astronomy \& Space Science, Kyung Hee University, Gyeonggi 446-701, Korea
\email{yy.choi@khu.ac.kr}
}

\date{Received September 15, 1996; accepted March 16, 1997}

 
\abstract
{}
{We study the environmental effects on the activity in galactic nuclei
  by comparing galaxies in clusters and in the field.}
{Using a spectroscopic sample of galaxies in Abell clusters 
  from the Sloan Digital Sky Survey Data Release 7,
  we investigate the dependence of nuclear activity
  on the physical parameters of clusters as well as the nearest neighbor galaxy.
We also compare galaxy properties between active galactic nuclei (AGN) hosts and non-AGN galaxies.}
{We find that the AGN fraction of early-type galaxies starts to decrease
  around one virial radius of clusters ($r_{\rm 200,cl}$) 
  as decreasing clustercentric radius,
  while that of late types starts to decrease
  close to the cluster center ($R\sim0.1-0.5 r_{\rm 200,cl}$).
The AGN fractions of early-type cluster galaxies, on average,
  are found to be lower than those of early-type field galaxies 
  by a factor $\sim 3$.
However, the mean AGN fractions of late-type cluster galaxies
  are similar to those of late-type field galaxies.
The AGN fraction of early-type brightest cluster galaxies
  lies between those of 
  other early-type, cluster and field galaxies with similar luminosities.
In the field, the AGN fraction is strongly dependent on the morphology of and 
  the distance to the nearest neighbor galaxy.
We find an anti-correlation 
  between the AGN fraction and the velocity dispersion of clusters
  for all subsamples divided by morphology and luminosity of host galaxies.
The AGN power indicated by $L_{\rm [OIII]}/M_{\rm BH}$
  is found to depend strongly on the mass of host galaxies
  rather than the clustercentric radius.
The difference in physical parameters
  such as luminosity, ($u-r$) colors, star formation rates, and ($g-i$) color gradients
  between AGN hosts and non-AGN galaxies is seen 
  for both early and late types at all clustercentric radii, while 
  the difference in structure parameters between the two
  is significant only for late types.
  }
{These results support the idea that
  the activity in galactic nuclei is triggered through galaxy-galaxy interactions and mergers
  when gas supply for AGN is available.}
{}


\authorrunning{H. S. Hwang et al.}
\maketitle
%

\section{Introduction}
What powers the activity in galactic nuclei?
It is generally accepted that active galactic nuclei (AGNs) are powered by accretion onto
  supermassive black holes (SMBHs, \citealt{lb69}), but
  it is still poorly understood what the source of fuel is
  and how the fuel can be accreted by removing its angular momentum 
  (see \citealt{jog06} for a review).
  
In the hierarchical picture of galaxy formation,
  massive galaxies are formed through accretion and mergers
  of other galaxies.
As the star formation (SF) is triggered through 
  galaxy-galaxy interactions and mergers
  by supplying gas into the center of galaxies,
  the activity in galactic nuclei is also expected to be triggered in a similar way
  with a gas inflow towards the center of galaxies by feeding SMBHs 
  \citep{san88,bh92,spr05,di05,hop06}.
On the other hand, 
  internal processes such as bar-driven gas inflow 
  (e.g., \citealt{com03}; see \citealt{kk04} for a review),
   turbulence in interstellar matter (e.g., \citealt{wada04}), and 
   stellar wind (e.g., \citealt{co07}) can also supply gas to SMBHs
   to trigger their activity.

Early studies of the environment of Seyfert galaxies
  have provided some hints for the triggering mechanism of the nuclear activity
  and its connection to neighbor galaxies
  (e.g., \citealt{pet82,dah84,dah85,keel85,fs88,vir00,ls95,dh99}).
As the Sloan Digital Sky Survey (SDSS; \citealt{york00})
  and Two Degree Field Galaxy Redshift Survey (2dFGRS; \citealt{col01})
  have produced unprecedented photometric and spectroscopic data of nearby galaxies,
  the physical properties of AGN host galaxies and their
  environmental dependence have been extensively studied, 
  providing strong observational
  constraints on the triggering mechanism of the activity in galactic nuclei
  (e.g., \citealt{mil03,kau03agn,kau04,wake04,best04,best05,ser06,sor06,cons06,cons08,
  ell08,li08agn,choi09,hag10,jhlee10env,sch10host,sch10early,pad10};
  see also \citealt{was05,sil09,mon09,cis11} for high-$z$ AGNs).
For example, 
  \citet{mil03} found no environmental dependence of the AGN fraction
  using the SDSS data.
Later, \citet{kau04} showed that 
  luminous AGNs with $L_{\rm [OIII]}>10^7$ ($L_\odot$)
  are dependent on the environment,
  in the sense that powerful AGNs
  are found predominantly in low-density regions.
In addition, 
  the fraction of galaxies having radio-loud AGNs with low emission-line luminosities
  is found to be high in high-density regions 
  compared to low-density regions \citep{best05}.
  
Regarding on the the activity in galactic nuclei and its connection to neighbor galaxies,
  some studies found no strong evidence 
  for the effects of galaxy-galaxy interactions and mergers 
  on the activity 
  \citep{derob98,mal98,sch01,grog05,pie07,li08agn,geo09,gab09,tal09,darg10,sm11},
  but other studies showed the evidence for the effects
  \citep{heck86,keel96,dom05,kou06agn,kuo08,urr08,com09,rog09,koss10,ell11,sil11}.
  
Focusing on this issue,
  our group has been investigating the dependence of the activity in galactic nuclei
  on host galaxy properties and environments.
For example, \citet{choi09} compared several physical parameters of
  AGN hosts and non-AGN galaxies using SDSS data release 5,
  and found that the AGN fraction depends 
  mainly on morphology and color of host galaxies.
Therefore, when morphology and color are fixed,
  the AGN fraction is almost independent of 
  luminosity or stellar velocity dispersion of host galaxies.
AGNs are found to be typically hosted by intermediate-color, late-type
  ($u-r=2- 2.4$) and bluish early-type galaxies 
  (peak at $u-r \sim 2.0$), 
  which indicates that AGN host galaxies have a
  younger stellar population than non-AGN galaxies at given
  luminosity or velocity dispersion.
Among the late-type galaxies,
  bluer color galaxies host more powerful AGNs.
These results support the idea that
  more massive and redder galaxies are unlikely to host AGNs
  because of lack of gas to feed their SMBHs.

We also studied the connection between the presence of bars and 
  the activity in galactic nuclei \citep{leegh11bar} using 
  the bar galaxy sample of \citet{leegh11samp},
  and found that the bar fraction in AGN-host galaxies is higher 
  than in non-AGN galaxies.
However, this trend is simply caused by the fact that AGN-host galaxies 
  are on average 
  more massive and redder than non-AGN galaxies since the bar fraction increases 
  with $u-r$ color and velocity dispersion. 
Therefore the excess of bar fraction in AGN-host
  galaxies disappears
  when AGN-host and non-AGN galaxies with fixed $u-r$ color and 
  velocity dispersion are compared.
These results suggest that the activity in galactic nuclei
  is not directly connected with the presence of bars.
  
As for the environmental dependence of nuclear activity,
  we found a strong dependence of the AGN fraction
  on the morphology of and the distance to the nearest neighbor galaxy (Choi et al. in prep.).
When an early-type galaxy has a close neighbor galaxy,
  the AGN fraction increases as it approaches a late-type neighbor, but
  decreases as it approaches an early-type neighbor.
For the late-type case,
  the AGN fraction also increases as it approaches a late-type neighbor, but
  remains constant or decreases as it approaches an early-type neighbor.
The bifurcations of the AGN fraction depending on the neighbor's morphology
    are seen around one virial radius of the neighbor.
This can support the idea of interaction-induced activity in galactic nuclei,
  and implies that hydrodynamic interactions
  with the nearest neighbor in addition to the tidal interactions
  play important roles in triggering nuclear activity.

One thing interesting in these studies of environmental dependence
  is a comparison of the activity in galactic nuclei in galaxy clusters 
  and in the field.
Since \citet{gis78} first found a lack of 
 optical emission-line galaxies in nearby clusters,
 many studies have extended this comparison of 
 the activity in galactic nuclei between cluster and field galaxies
 using an extensive survey data sets in the local universe
 (e.g., \citealt{dre85, pb06, best07, arn09, von10, kp10}).
For example,
  \citet{arn09} confirmed an increasing AGN fraction
  from cluster to group regions.
One of the most interesting results in their study
  is that this trend remains the same even if they fix 
  the morphology of host galaxies (e.g. early-type galaxy).
This suggests that
  the change in AGN fraction is not simply caused by 
  the morphological mix with the environment,
  but is directly connected to environment.
  
Thanks to recent, powerful X-ray observatories 
  such as {\it XMM-Newton} and {\it Chandra}, 
  this comparison has been extended to higher redshifts
  so that one can study the evolution of AGNs as a function of redshift 
  (e.g., \citealt{mar06, mar09, john03, re05, east07, gil09}).
For example, 
 using the cluster galaxies at $z\sim0.05-1.3$,
  \citet{mar09} found an increase of AGN fraction
  by a factor of eight at $z=1$ compared to the local universe
  with 3.8$\sigma$ statistical significance.
This strong evolution of AGN fraction is qualitatively comparable to
  the evolution of SF galaxy fraction (the Butcher-Oemler effect; \citealt{bo84}),
  which suggests a close connection between SF and nuclear activity
  (e.g., \citealt{mul11}).
  
It should be noted that the statistics can vary depending on 
  the AGN selection criteria 
 (i.e. optical line ratio, X-ray luminosity, mid-infrared color, 
  or radio luminosity).
For example, \citet{gal09} constructed
  AGN samples using three different selection criteria 
  (mid-IR color, radio luminosity, and X-ray luminosity),
  and found that
  the increase of cluster AGN surface density with redshift
  on average is more steep than that of field quasars.  
Although they found significant differences 
  between the AGN populations identified by different methods,
  their results do not change much depending on the selection criteria
 (see also \citealt{atl11}).

On the other hand, 
  it was found that galaxy properties such as morphology, luminosity, and
  star formation rate (SFR) strongly depend on the distance to and 
  the morphology of the nearest
  neighbor galaxy, which indicates an important role of hydrodynamic interactions 
  with neighbor galaxies \citep{park08,pc09,hp09,hp10,hwa10lirg,hwa11inter}.
This dependence was found even when the large-scale background density is fixed, 
  and thus is completely different from the commonly 
  known morphology $-$ local density relation.
We also found that successive interactions with nearby galaxies are still important
  in determining the morphology or SFR even in galaxy cluster regions \citep{ph09,cer11}.
Since the activity in galactic nuclei is also expected to be triggered through
  galaxy-galaxy interactions and mergers as we discussed, it is necessary to investigate
  the role of neighbor galaxies in determining the activity
  hoping that the difference of nuclear activity between cluster and field galaxies
  be understood.
  
In this paper, we study the the activity in galactic nuclei of cluster and field galaxies 
  using the SDSS data,
  and compare physical parameters of AGN hosts and non-AGN galaxies.
Section \ref{data} describes the data used in this study.
Environmental dependence of the activity in galactic nuclei
  and the comparison of galaxy properties between AGN hosts and non-AGN galaxies
  are given in \S \ref{results}.
Discussion and conclusions are given in \S \ref{discuss} and \S \ref{con}, respectively.
Throughout this paper, we adopt a flat $\Lambda$CDM cosmological model with density parameters 
  $\Omega_{\Lambda}=0.73$ and $\Omega_{m}=0.27$.


\section{Observational data set}\label{data}
\subsection{Sloan Digital Sky Survey sample}\label{sdss}

We used a spectroscopic sample of galaxies
  including the main galaxy sample ($m_r<17.77$) and 
  faint galaxies ($m_r>17.77$) whose spectroscopic redshifts are available 
  in the SDSS data release 7 (DR7, \citealt{aba09}),
  which results in 915,327 galaxies.
Completeness of the spectroscopic data in SDSS is poor
  for bright galaxies with $m_r<14.5$ because of
  the problems of saturation and cross-talk in the spectrograph, and
  for the galaxies located in high-density regions such as galaxy clusters
  due to the fiber collision.
Thus, we added a photometric sample of galaxies with $m_r<17.77$
  whose redshift information is available in the literature 
 (see \citealt{hwa10lirg} for more detail).
In total, the redshift information for 186,055 galaxies in the photometric sample
  was compiled, which overlaps with 174,634 galaxies in the spectroscopic sample. 
Finally, we added 11,421 galaxies to the spectroscopic sample of 
  915,327 galaxies,
  which yields a final sample of 926,748 galaxies.
In the result, the spectroscopic completeness of our sample
  is higher than $85\%$ at all magnitudes with $m_r<17.77$ 
  and even in the center of galaxy clusters (see Fig. 1 in \citealt{ph09}).
Note that the final sample also contains the galaxies 
  whose spectral classifications provided in the SDSS are QSOs.

To investigate the physical parameters of galaxies,
  we used several value-added galaxy catalogs (VAGCs) drawn from SDSS data.
Photometric and structure parameters were adopted from SDSS pipeline \citep{sto02},
  and spectroscopic parameters were from 
  the MPA/JHU DR7 VAGC\footnote{http://www.mpa-garching.mpg.de/SDSS/DR7/}.  
We adopted the galaxy morphology information from
  Korea Institute for Advanced Study (KIAS) DR7 VAGC\footnote{http://astro.kias.re.kr/vagc/dr7/} 
  \citep{pc05,choi10},
  which contains 697,320 main galaxy sample in NYU VAGCs as well as
  10,497 photometric sample of galaxies with the redshift information 
  from various existing redshift catalogs.
We performed additional visual classification 
  for the galaxies in DR7 that are not included in KIAS DR7 VAGC.
During the visual inspection of color images of galaxies,
  we eliminated 1649 spurious sources
  (e.g., faint fragments of bright galaxies, 
  diffraction spikes of bright stars), and
  they are not included in the final sample of 926,748 galaxies.

\subsection{Cluster Sample and Galaxy Membership in Clusters}\label{membership}

The galaxy and cluster samples used in this study are similar to
  those in \citet{ph09}, but for SDSS DR7 data.
We used the Abell catalog of galaxy clusters \citep{aco89} to
  identify cluster galaxies in our galaxy sample.
Among the Abell clusters,
  we selected those that have known spectroscopic
  redshifts in the NED.
We found $910$ clusters located within the SDSS survey region.
We adopted the position of cluster center in the NED,
  and revised it when the center position determined in X-ray is available in the literature.

To determine the membership of galaxies in a cluster,
  we used the ``shifting gapper'' method of \citet{fad96}.
In the radial velocity versus clustercentric distance space, 
  the cluster member galaxies are selected by grouping galaxies
  with connection lengths of 950 km s$^{-1}$ in the direction of the radial velocity
  and of 0.1 $h^{-1}$Mpc in the direction of the clustercentric radius $R$.
Grouping in the radial velocity direction is made within each distance bin
  with 0.2 $h^{-1}$Mpc width.
A larger bin width is used when the number of galaxies in a bin is less than 15.
If the boundary is not reached out to $R=3.5 h^{-1}$Mpc, we stopped
  the grouping at $R=3.5h^{-1}$Mpc.
We iterate the procedure until the number of cluster members has converged.
From this procedure we
  obtained 240 Abell clusters which have more than or equal to 10 member galaxies.

We computed a radius of $r_{200,{\rm cl}}$ (usually called the virial radius)
  for each cluster where the mean overdensity drops to 200
  times the critical density of the universe $\rho_{\rm c}$,
  using the formula given by \citet{car97}:
\begin{equation}
r_{200,{\rm cl}}= \frac{3^{1/2}\sigma_{\rm cl}}{10 H(z)},
\end{equation}
where $\sigma_{\rm cl}$ is a velocity dispersion of the cluster and
  the Hubble parameter at $z$ is
  $H^2(z)=H^2_0 [\Omega_m(1+z)^3 +\Omega_k(1+z)^2+\Omega_\Lambda]$ \citep{pee93}.
$\Omega_m$, $\Omega_k$, and $\Omega_\Lambda$ are the dimensionless density parameters.

The velocity dispersion was computed for each cluster,
  using the galaxies that are identified as members in the cluster main body
  by rejecting interlopers.
To reject the interlopers,
  we computed $\delta$, which indicates
  the local deviation of the radial velocity of a galaxy
  from the systemic velocity ($v_{\rm sys}$) of the entire cluster
  taking into account the velocity dispersion ($\sigma_{\rm cl,all}$) of the cluster.
It is defined by
\begin{equation}
\delta^2 =  \frac{N_{nn}}{\sigma_{\rm cl,all}^2} \left[ (v_{\rm local}-v_{\rm sys})^2 +
(\sigma_{\rm local}-\sigma_{\rm cl,all})^2 \right],
\end{equation}
where $N_{nn}$ is the number of galaxies that defines the local environment, taken
  to be ${N_{\rm gal}}^{1/2}$ in this study.
We then used the galaxies with $\delta\leq3.0$
  to calculate the cluster velocity dispersion.

In addition to the sample of cluster galaxies obtained 
  adopting the ``shifting gapper'' method above,
  we included the galaxies located at projected separations of $R_{max}<R<10r_{\rm 200,cl}$ 
  to investigate the variation of galaxy properties
  over a wide range of clustercentric radius.
$R_{max}$ is the largest clustercentric distance of the cluster member galaxies determined above,
  which is typically $\sim1.3~h^{-1}$Mpc.
These additional galaxies are constrained to have velocity difference
  relative to the cluster's systematic velocity
  less than $\Delta v=|v_{gal}-v_{sys}|=1000$ km s$^{-1}$.
The final sample consists of galaxies smoothly distributed 
  from the cluster center to $R=10r_{\rm 200,cl}$ for each cluster.

\begin{figure}
\center
\includegraphics[scale=0.4]{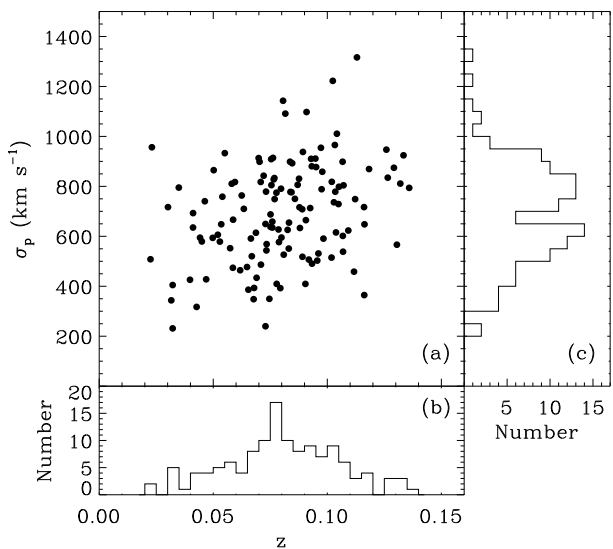}
\caption{Distribution of redshifts and velocity dispersions
  for our 129 relaxed Abell clusters.
}\label{fig-cl}
\end{figure}

We rejected the clusters that appear to be interacting or merging,
  which are found in the galaxy velocity versus clustercentric distance space.
Dynamically young clusters 
  with the brightest cluster galaxy (BCG) at large clustercentric distance
  ($R_{\rm BCG}>0.5r_{\rm 200,cl}$) were also rejected.
These procedures left us
  only dynamically relaxed clusters,
  so our results are not affected by 
  violent SF or nuclear activity of cluster galaxies
  caused by cluster interactions or mergers (e.g., \citealt{hl09}).

We included eight clusters to increase the statistics,
   which were originally eliminated in previous studies 
   due to the incomplete survey coverage out to 10$r_{\rm 200,cl}$
   \citep{hwa10lirg,cer11}.
We finally obtain a sample of 129 {\it relaxed} Abell clusters
 with $0.02\lesssim z\lesssim 0.14$ and 
 $200\lesssim \sigma_{\rm p} \lesssim 1350$ (km s$^{-1}$),
 and present the distribution of redshifts and 
 velocity dispersions for these clusters
 in Fig. \ref{fig-cl}.

\subsection{AGN Selection}\label{agnsel}

\begin{figure}
\center
\includegraphics[scale=0.35]{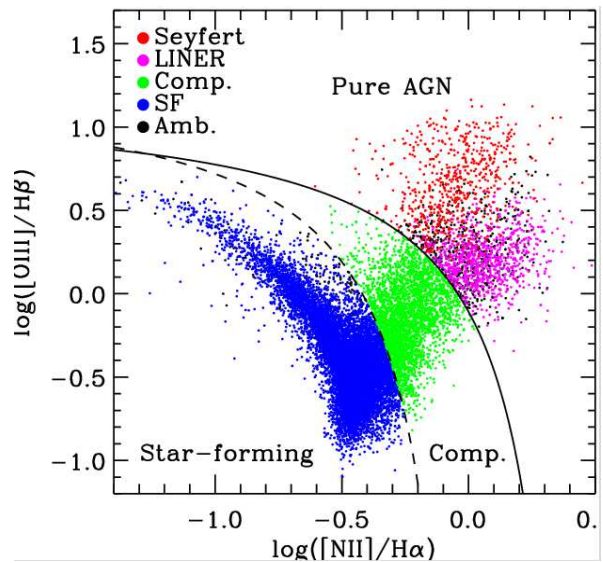}
\caption{[OIII]/H$\beta$ vs. [NII]/H$\alpha$ line ratio diagram for our sample galaxies.
Different spectral types following the scheme of \citet{kew06} 
  are represented by different colored symbols
  (Seyfert: red, LINER: pink, Composite: green, SF: blue, Ambiguous: black).
The solid and dashed lines indicate the extreme starburst \citep{kew01}
  and pure SF limits \citep{kau03agn}, respectively.
}\label{fig-bpt}
\end{figure}

We determined the spectral types of emission-line galaxies 
  based on the criteria given by \citet{kew06} 
  using the emission line ratio diagrams, 
  commonly known as the Baldwin-Phillips-Terlevich (BPT) diagrams \citep{bpt81,vo87}. 
In brief, for the galaxies with signal-to-noise ratio (S/N)$\geq$3
   in the strong emission-lines H$\beta$, [OIII] $\lambda$5007,
   H$\alpha$, [NII] $\lambda$6584, and [SII]$\lambda\lambda$6717,6731,
   we determined their spectral types based on their positions 
   in the line ratio diagrams
  on which [OIII]/H$\beta$ is plotted against
   [NII]/H$\alpha$, [SII]/H$\alpha$, and [OI]/H$\alpha$ :
    star-forming galaxies, Seyferts, 
    low-ionization nuclear emission-line regions (LINERs), 
    composite galaxies, and ambiguous galaxies.
Composite galaxies host a mixture of star-formation and AGN,
  and lie between the extreme starburst line \citep{kew01} and the
  pure star formation line \citep{kau03agn}
  in the [OIII]/H$\beta$ vs. [NII]/H$\alpha$ line ratio diagram (see Fig. \ref{fig-bpt}).
Ambiguous galaxies are those classified as one
  type in one or two diagrams, but as another type
  in the other diagrams (see \citealt{kew06} for more detail).
LINERs are generally thought to be low-luminosity AGNs,
  but because of other mechanisms that can produce the LINER-like spectra,
  the AGN excitation mechanism of LINERs 
  is still debated (see \citealt{ho08} for a review).
However, from a recent X-ray analysis of 82 LINERs, 
  \citet{gm09} concluded that 
  the observational data support the AGN nature for 80\% of their LINER sample 
  (but see also \citealt{sar10}), which suggests that
  the contamination of non-AGN LINERs to our analysis (which is based on total AGN samples)
  is not significant.

\begin{figure}
\center
\includegraphics[scale=0.75]{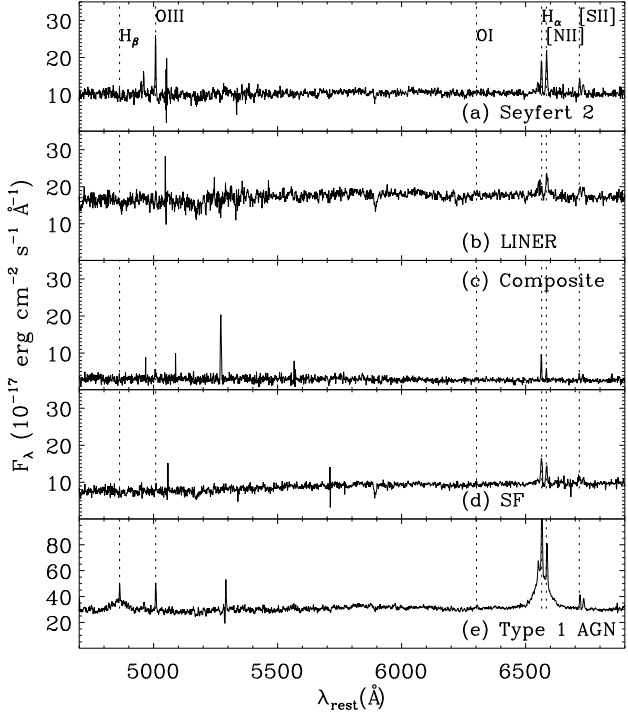}
\caption{Example SDSS spectra with low S/N (S/N$_{\rm H\alpha} \approx 5-7$) of several spectral types:
  (a) Seyfert 2 (SDSS ObjId: 588007006326096041), 
  (b) LINER (ObjId: 587736783609069705), 
  (c) Composite (ObjId: 587742061625606544), 
  (d) SF (ObjId: 587728677931778286), and 
  (e) Type 1 AGN (ObjId: 587727225695698974).
Vertical dashed lines indicate the positions of 
  emission lines used for the spectral classification.
}\label{fig-spec}
\end{figure}

We assign `undetermined' type to those that do not satisfy S/N criteria.
We restrict our analysis to the galaxies at $z>0.04$ 
  owing to the problem of small (3\arcsec) fixed-size aperture \citep{kew06}.
In Fig. \ref{fig-spec}, we show example spectra of
  several spectral types, especially for low S/N cases (S/N$_{H\alpha}\approx5-7$).

These AGN criteria select only Type II AGNs with narrow emission lines, 
  and miss Type I AGNs with broad Balmer lines.
To secure Type I AGNs missed in this method,
  we included galaxies whose spectral classifications (\texttt{specClass}) 
  provided by SDSS pipeline are quasar 
  (i.e. \texttt{specClass} = \texttt{SPEC}$\_$\texttt{QSO} or
  \texttt{SPEC}$\_$\texttt{HIZ}$\_$\texttt{QSO}; 
  see \citealt{sto02} for more detail).
Among 472 370 spectroscopic sample of galaxies at $0.04\leq z<0.1434$ 
  that we are interested in, 
  2206 quasars ($\sim0.5\%$) with broad emission lines were included.
The statistics for AGN subsamples are summarized in Table \ref{tab-agn},
  which lists the fraction and the number of each subsample.
  
In this study, the AGN fraction means a ratio of 
  the number of Type I plus Type II AGNs 
  (Seyferts, LINERs and composite galaxies determined in the line ratio diagram)
  to the number of galaxies with spectroscopic parameters measured in MPA/JHU DR7 VAGC.
If we use more strict AGN definition by excluding composite galaxies,
  our conclusions do not change but our statistics are worse because of smaller number of galaxies.
Note also that the photometric sample of galaxies with redshift adopted from the literature
  is not included when we compute the AGN fraction because their
  spectral types can not be determined.
Some previous studies suggested that the environments 
  attributed to companion galaxies are
  different between Seyfert 1 and 2 (e.g., \citealt{lau94,kou06agn}).
In addition, LINERs are known to be phenomenologically different from Seyferts 
  (see \citealt{ho08} for a review).
Therefore, it should be noted that
  our analysis based on AGNs including all subsamples
  does take into account the difference between the subsamples, 
  and suggests only a broad consensus on the activity of galactic nuclei.
A detailed analysis focusing on the difference between the subsamples
  needs to be conducted with a more
  extensive data set in future studies.

\begin{table*}
\begin{center}
\caption{Sample statistics\label{tab-agn}}
\begin{tabular}{ccrrrrrrrr}
\hline\hline 
 Sample & Morph.$^{a}$  & Seyferts & LINERs & Composite & Type I  &  SF    & Ambiguous  & Undetermined & Total \\
\hline
     C1 & ETGs &   1.4\%(014) &   1.5\%(016) &   6.1\%(0063) &   0.0\%(00) &   6.7\%(0069) &   1.1\%(011) &  83.3\%(0863) & 100\%(1036) \\
        & STGs &   0.6\%(009) &   0.3\%(005) &   6.8\%(0107) &   0.0\%(00) &  72.2\%(1132) &   2.0\%(031) &  18.1\%(0284) & 100\%(1568) \\
     C2 & ETGs &   1.3\%(099) &   2.5\%(195) &   4.8\%(0379) &   0.0\%(02) &   2.3\%(0179) &   1.5\%(118) &  87.7\%(6921) & 100\%(7893) \\
        & STGs &   1.7\%(150) &   1.2\%(102) &  12.5\%(1085) &   0.3\%(29) &  51.6\%(4465) &   1.5\%(134) &  31.1\%(2695) & 100\%(8660) \\
     C3 & ETGs &   0.8\%(057) &   5.5\%(369) &   3.6\%(0244) &   0.0\%(03) &   0.8\%(0053) &   2.0\%(136) &  87.2\%(5858) & 100\%(6720) \\
        & STGs &   4.4\%(200) &   5.2\%(238) &  17.7\%(0809) &   1.6\%(73) &  25.8\%(1179) &   3.9\%(178) &  41.5\%(1901) & 100\%(4578) \\
     \hline\hline
\end{tabular}
\begin{flushleft}
$^{a}$: Galaxy morphology (ETGs : early-type galaxies, LTGs : late-type galaxies).
\end{flushleft}
\end{center}
\end{table*}

\subsection{Physical parameters of galaxies}

The physical parameters of galaxies that we consider in this study are
  $r$-band absolute Petrosian magnitude ($M_r$),
  morphology, axis ratio,
  ($u-r$) color, SFR, [OIII] emission line flux,
  ($g-i$) color gradient,
  concentration index ($c_{\rm in}$),
  internal velocity dispersion ($\sigma$), and
  Petrosian radius in $i$-band.
Here we give a brief description of these parameters.

The $r$-band absolute magnitude $M_r$ was computed using the formula,
\begin{equation}\label{eq-mag}
M_r=m_r-DM-K(z)+E(z),
\end{equation}
where $DM$ is a distance modulus, 
$K(z)$ is the $K$-correction, and $E(z)$ is the luminosity evolution correction.
$DM$ is defined by $DM\equiv5 {\rm log}(D_L/10)$ and 
  $D_L$ is a luminosity distance in unit of pc.
The rest-frame absolute magnitudes of
  individual galaxies are computed in fixed bandpasses, shifted to $z=0.1$,
  using Galactic reddening correction \citep{sch98} and $K$-corrections
  as described by \citet{bla07kcorr}.
The evolution correction given by \citet{teg04}, $E(z) = 1.6(z-0.1)$, is also applied.

\begin{figure}
\center
\includegraphics[scale=0.4]{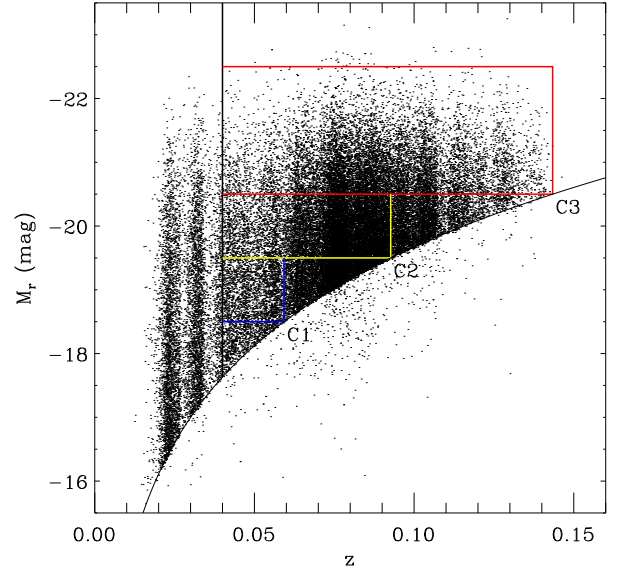}
\caption{Sample definitions of our three volume-limited samples
  in the absolute magnitude vs. redshift space.
The bottom curve indicates the apparent magnitude limit ($m_r=17.77$)
  for the main galaxy sample in SDSS 
  using the mean $K$-correction relation given by equation (2) of \citet{choi07}.
}\label{fig-vol}
\end{figure}

Figure \ref{fig-vol} shows the $r$-band absolute magnitudes of the cluster galaxies
  against their redshifts.
We define three volume-limited samples of galaxies
  using the redshift and absolute magnitude conditions as follows:
  C1 (faint galaxies: $-18.5\geq M_r>-19.5$ and $0.04\leq z\leq0.0593$),
  C2 (intermediate-luminosity galaxies: $-19.5\geq M_r>-20.5$ and $0.04\leq z\leq0.0927$), and
  C3 (bright galaxies: $-20.5\geq M_r>-22.5$ and $0.04\leq z\leq0.1434$).

The $^{0.1}(u-r)$ color was computed with the extinction and $K$-corrected model magnitudes.
The superscript 0.1 means the rest-frame magnitude $K$-corrected to the redshift of 0.1,
  and will subsequently be dropped.

We adopt the values of $(g-i)$ color gradient, concentration index ($c_{\rm in}$),
  and Petrosian radius $R_{\rm Pet}$
  computed for the galaxies in KIAS VAGC \citep{choi10}.
The $(g-i)$ color gradient was defined by the color difference
  between the region with $R<0.5R_{\rm Pet}$ and the annulus with
  $0.5R_{\rm Pet}<R<R_{\rm Pet}$,
  where $R_{\rm Pet}$ is the Petrosian radius estimated in $i$-band image.
To account for the effect of flattening or inclination of galaxies,
  elliptical annuli were used to calculate the parameters.
The (inverse) concentration index is defined by $R_{50}/R_{90}$,
  where $R_{50}$ and $R_{90}$ are semimajor axis lengths of ellipses
  containing $50\%$ and $90\%$ of the Petrosian flux in the $i$-band image, respectively.

The velocity dispersion of galaxies is adopted from
  the measurements by an automated
  spectroscopic pipeline called \texttt{specBS}, 
  which was written by D. J. Schlegel (in prep.).
We performed the aperture correction following \citet{cap06},
\begin{equation}
\sigma_{\rm corr} =  \left(\frac{R_{\rm fib}}{R_{\rm eff}}\right)^{0.066\pm0.035} \sigma_{\rm fib},
\end{equation}
where
$\sigma_{\rm fib}$ is the measured velocity dispersion from the SDSS spectra 
  obtained by an optical fiber with radius of $R_{\rm fib}=1.5\arcsec$.
$R_{\rm eff}$ is an effective radius, which is derived by 
  $R_{\rm eff}=(b/a)^{0.5}_{\rm deV}r_{\rm deV}$,
  where $r_{\rm deV}$ is the seeing-corrected effective radius 
  along the major axis of the galaxy
  derived from a model fit of the de Vaucouleurs profile in the $i$-band,
  and $b/a$ is an axis ratio (minor to major) of de Vaucouleurs fit.
For statistical analysis in \S \ref{results}, 
  we do not use galaxies with values less than 
  the instrumental resolution ($\sim 70$ \kms).
  
The black hole mass ($M_{\rm BH}$) is computed 
  through the $M_{\rm BH}-\sigma$ relation,
\begin{equation}
{\rm log}~(M_{\rm BH}/M_\odot) = \alpha +\beta {\rm log}(\sigma/200~{\rm km~s^{-1}}).
\end{equation}
We adopt $\alpha$ and $\beta$ values differently depending on the morphology of host galaxies
 (\citealt{gul09}; $\alpha=8.22\pm 0.073$ and $\beta=3.86\pm0.380$ for early types,
  and $\alpha=7.95\pm 0.286$ and $\beta=4.58\pm1.583$ for late types).
$M_{\rm BH}$ is computed for only galaxies with velocity dispersion values 
  larger the instrumental resolution ($\sim 70$ \kms).

\begin{figure*}
\center
\includegraphics[scale=0.62]{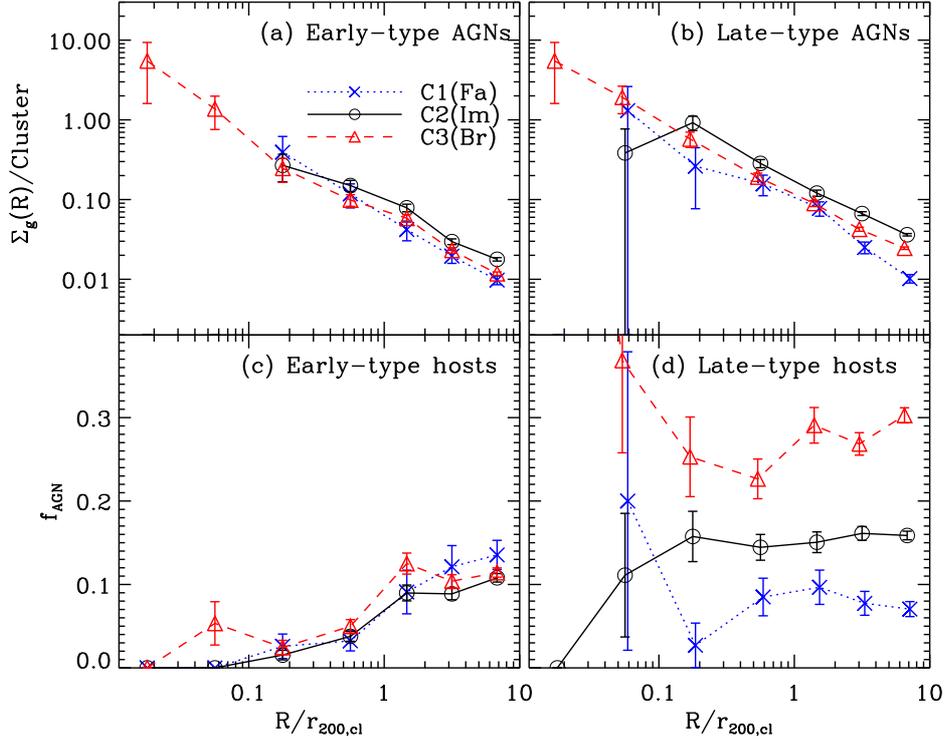}
\caption{
({\it top}) Projected number density of AGN host galaxies for (a) early and (b) late types
  as a function of
  the clustercentric radius normalized to the cluster virial radius $R/r_{\rm 200,cl}$.
({\it bottom}) AGN fraction ($f_{\rm AGN}$) as a function of
  the clustercentric radius for (c) early- and (d) late-type galaxies.
Error bar indicates Poissonian uncertainty.
}\label{fig-fagn1d}
\end{figure*}

The [OIII] emission line fluxes are taken from MPA/JHU DR7 VAGC \citep{tre04},
  which were computed using the straight integration over the fixed bandpass
  from the continuum-subtracted emission line.
We corrected the line fluxes for internal extinction using the Balmer decrement and 
  the reddening curve assuming an intrinsic H$\alpha$/H$\beta$
  flux ratio of 2.85 for star-forming galaxies and 3.1 for AGN host galaxies 
  (the Balmer decrement for
  case B recombination at T = 10$^4$ K and N$_e\sim10^1-10^4$ cm$^{-3}$; \citealt{ost06}).
For [OIII] line luminosity ($L_{\rm [OIII]}$) measurements, 
  we use only the galaxies with S/N$_{\rm [OIII]}>3$.
We adopt $L_{\rm [OIII]}$ as an accretion rate indicator 
  (\citealt{kau03agn,heck05}, but see also \citealt{tb10}),
  then use $L_{\rm [OIII]}/M_{\rm BH}$ that is proportional to the Eddington ratio
  as an indicator of AGN power.

The SFRs of galaxies are also adopted from
  the MPA/JHU DR7 VAGC \citep{bri04},
  which provides extinction and aperture corrected
  SFR estimates of star-forming galaxies
  as well as other types of galaxies 
  (e.g., AGN, Composite, low S/N SF, low S/N LINER, and unclassifiable).
For those galaxies where they can not directly
  measure SFRs from the emission lines
  such as AGN and composite galaxies,
  they use the 4000-$\AA$ break (D4000) to estimate SFRs
  (see \citealt{bri04} and 
  http://www.mpa-garching.mpg.de/SDSS/DR7/sfrs.html for more detail).
We also use stellar mass estimates from the MPA/JHU DR7 VAGC,
  which are based on the fit of SDSS five-band photometry 
  with the model of \citet{bc03} (see also \citealt{kau03}).
We convert SFR and stellar mass estimates in MPA/JHU DR7 VAGC
  that are based on Kroupa IMF \citep{kro01}
  to those with Salpeter IMF \citep{sal55} by dividing them by a factor of 0.7 \citep{elb07}.

In our analysis we often limit the late-type galaxy sample to galaxies with
  $i$-band isophotal axis ratio $b/a$ greater than 0.6 
  (e.g., Figs \ref{fig-mag1d}-\ref{fig-sf1d}).
This is to reduce the effects of internal extinction on our results.
The absolute magnitude and color of late-type galaxies 
  with $b/a < 0.6$ can be inaccurate
  (see Figs. 5 and 12 of \citealt{choi07}).
Therefore, including them in the analysis may introduce
  a large dispersion in luminosity, color, or color gradient.

\subsection{Nearest neighbor galaxy in clusters}

To account for the effects of the nearest neighbor galaxy in cluster environment,
  we determine the distance and the morphology of the nearest neighbor galaxy.

We define the nearest neighbor galaxy of a target galaxy with absolute magnitude $M_r$
  as the one which
    is located closest to the galaxy on the sky and
    is brighter than $M_r+\Delta M_r$ among those in our cluster galaxy sample.
We adopt $\Delta M_r =0.5$.
We do not use the velocity condition to determine the nearest neighbor galaxy
  because it is selected from the cluster galaxy sample
  for which the velocity condition is already applied.

We obtain the nearest neighbor distance normalized by the virial radius 
  of the nearest neighbor as follows.
The virial radius of a galaxy within which 
  the mean mass density is 200 times the critical density 
  of the universe ($\rho_c$) is computed by
\begin{equation}
r_{\rm vir} = (3 \gamma L  / 4\pi / 200{\rho}_c )^{1/3},
\label{eq-vir}
\end{equation}

where $L$ is the galaxy luminosity, and $\gamma$ is the mass-to-light ratio.
We assume that the mass-to-light ratio of early-type galaxies
  is on average twice as large as that of late-type galaxies
  at the same absolute magnitude $M_r$,
  which means $\gamma$(early)$=2\gamma$(late) 
  [see \S 2.5 of \citet{pc09} and \S 2 of \citet{park08}]. 

Since we adopt $\Omega_m = 0.27$, $200\rho_c = 200 {\bar\rho}/\Omega_m = 740{\bar\rho}$
  where $\bar\rho$ is the mean density of the universe.
The value of mean density of the universe, 
  $\bar\rho=(0.0223\pm0.0005)(\gamma L)_{-20} (h^{-1}{\rm Mpc})^{-3}$,
  was adopted where $(\gamma L)_{-20}$ is the mass of a late-type galaxy 
  with $M_r=-20$ \citep{park08}.
According to our formula the virial radii of galaxies with
  $M_r=-19.5,-20.0,$ and $-20.5$ are 260, 300, and 350 $h^{-1}$ kpc for early types,
  and 210, 240, and 280 $h^{-1}$ kpc for late types, respectively.

\begin{figure*}
\center
\includegraphics[scale=0.6]{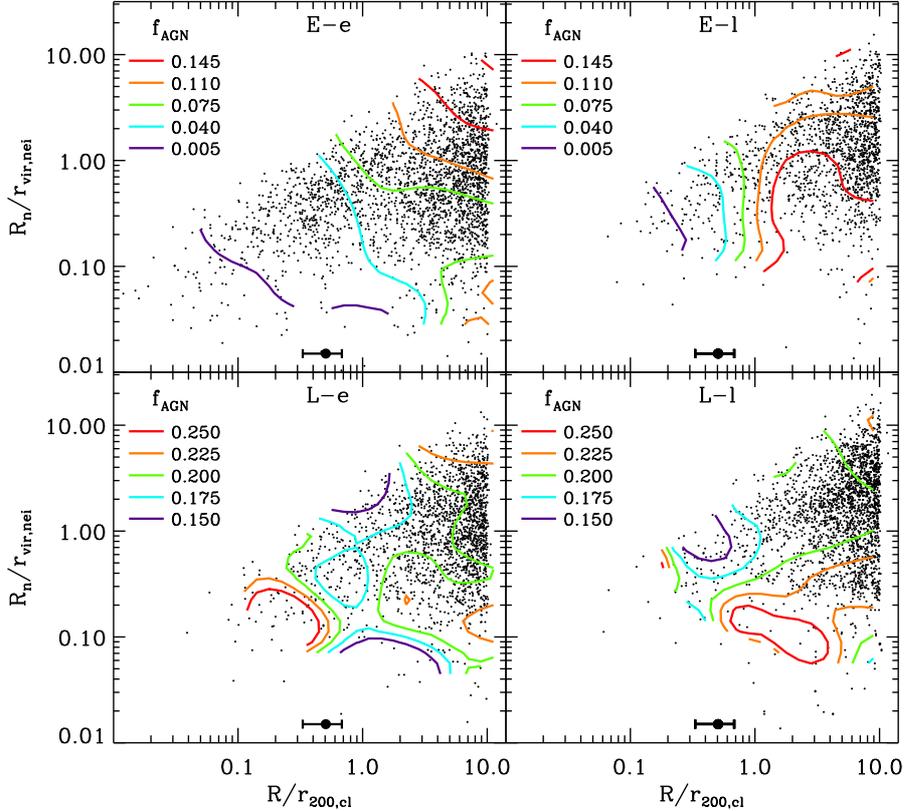}
\caption{AGN fraction ($f_{\rm AGN}$) contours 
  in the projected pair separation $R_n/r_{\rm vir,nei}$
  vs. the clustercentric distance $R/r_{\rm 200,cl}$ for the galaxies with 
  $0.04\leq z<0.0927$ and $-20.0\geq M_r>-20.5$.  
Four cases are given;
  the early-type target galaxies having an early-type neighbor (E-e),
  the early-type target galaxies having a late-type neighbor (E-l),
  the late-type target galaxies having an early-type neighbor (L-e), and
  the late-type target galaxies having a late-type neighbor (L-l).
The points with error bars above the x-axes denote the average 
 virial radius of BCGs.
}\label{fig-fagn2d}
\end{figure*}

\section{Results}\label{results}
\subsection{Activity in galactic nuclei of cluster and field galaxies}\label{env}

Figure \ref{fig-fagn1d} shows the projected number density of AGN host galaxies
  and the AGN fraction as a function of a projected clustercentric radius $R$
  normalized by the cluster virial radius $r_{\rm 200,cl}$.
The AGN fraction is the ratio of 
  the number of Type I plus Type II AGNs 
  (Seyferts, LINERs and composite galaxies determined in the line ratio diagram)
  to the number of galaxies with spectroscopic parameters measured in MPA/JHU DR7 VAGC.
They are shown for three luminosity ranges.
It shows that the projected number density of AGN hosts
  keeps increasing as the distance to the center of galaxy clusters decreases.
The slopes for three luminosity samples are not significantly different.

\begin{figure*}
\center
\includegraphics[scale=0.6]{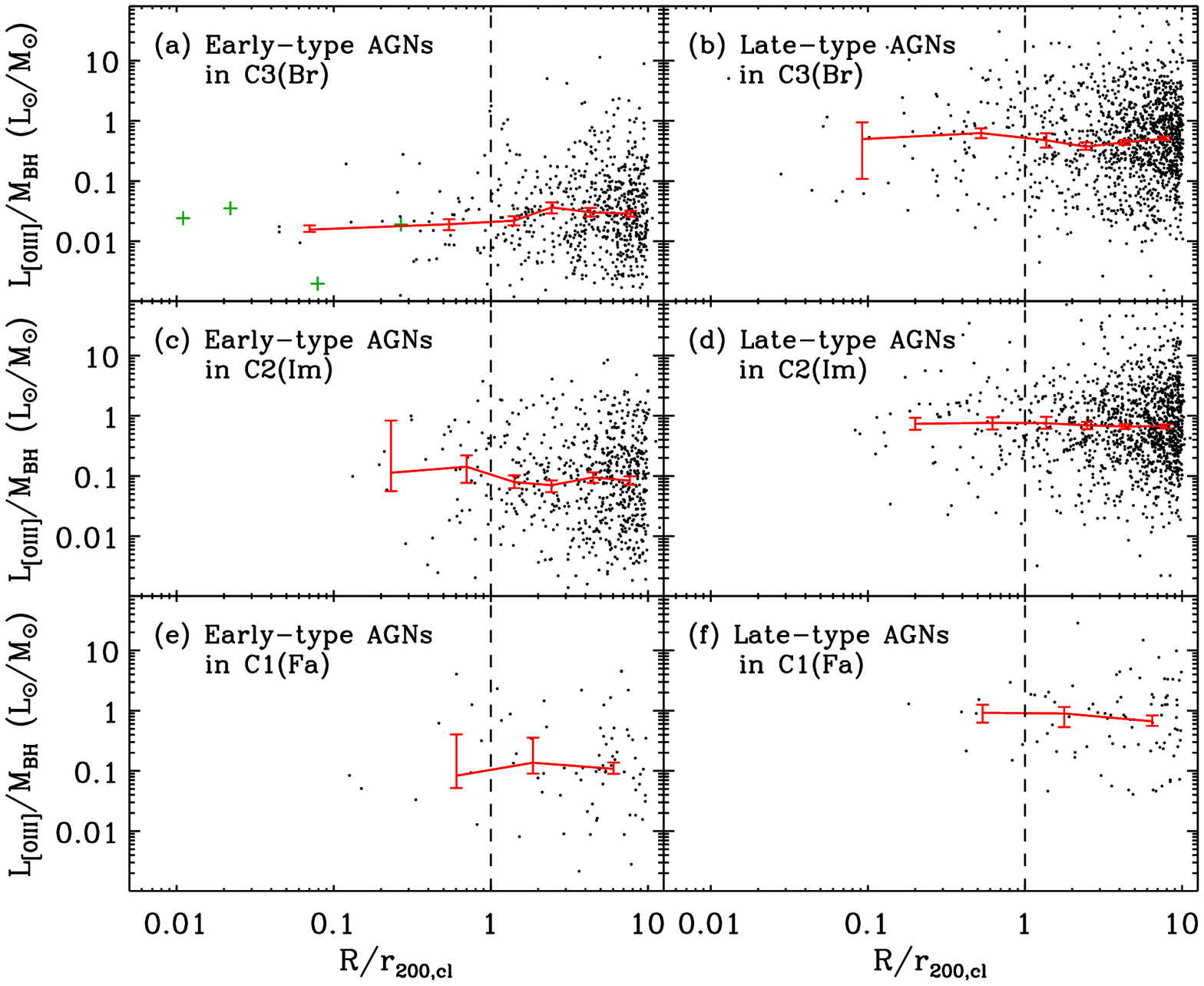}
\caption{$L_{\rm [OIII]}/M_{\rm BH}$ as a function of
  the clustercentric radius
  for ({\it left}) early- and ({\it right}) late-type galaxies.
Solid line is the median curve of each sample [C1({\it bottom})$-$C3({\it top})].
The BCGs, marked as crosses in top left panel, 
  are not used in calculating median values.
$L_{\rm [OIII]}$ is computed with $H_0 = 70$ km s$^{-1}$ Mpc$^{-1}$.
}\label{fig-ed1d}
\end{figure*}

\begin{figure*}
\center
\includegraphics[scale=0.6]{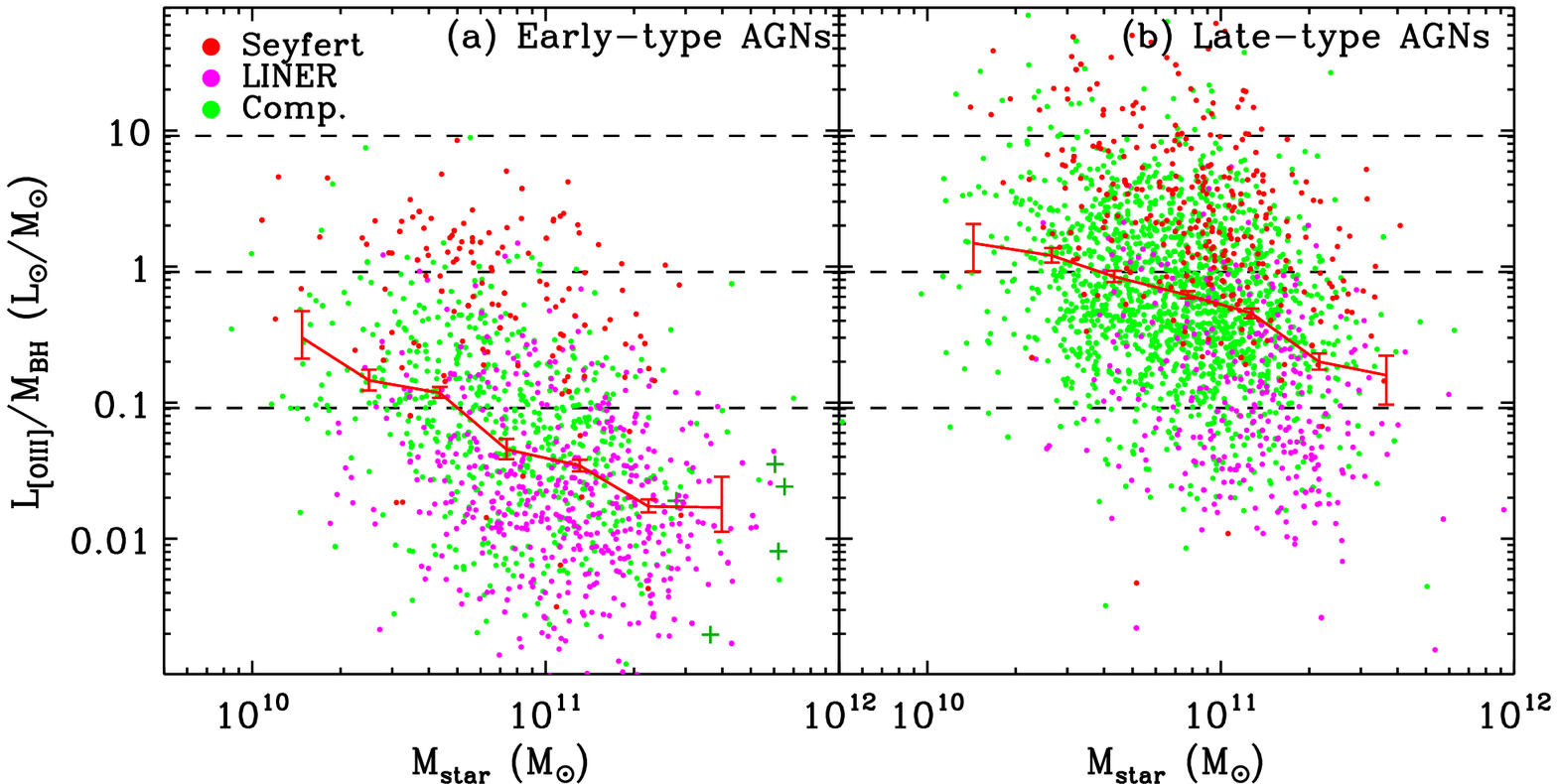}
\caption{$L_{\rm [OIII]}/M_{\rm BH}$ vs. stellar masses
  for ({\it left}) early- and ({\it right}) late-type galaxies.
Solid line is the median curve of each sample.
Seyfert, LINERs, and composite galaxies are
  denoted by red, pink, and green colored symbols.
Dashed lines from top to bottom 
  indicate the constant $L_{\rm bol}/L_{\rm edd}=$1, 0.1, and 0.01
  by assuming a bolometric correction 3500 for $L_{\rm [OIII]}$ \citep{heck04}.
The BCGs, marked as crosses,
  are not used in calculating median values.
$L_{\rm [OIII]}$ and $M_{\rm star}$ are computed with $H_0 = 70$ km s$^{-1}$ Mpc$^{-1}$.
}\label{fig-edm}
\end{figure*}
  
On the other hand,
  the AGN fraction (f$_{\rm AGN}$) for early types is almost constant 
  at large distance ($>r_{\rm 200,cl}$), 
  and starts to decrease inward around one virial radius of cluster.
This is consistent with Fig. 12 in \citet{von10},
  which shows a gradual decline of the AGN fraction in red galaxies
  at $R<r_{\rm 200,cl}$.
The late-type galaxies show a similar trend, but the fraction starts to drop
  closer to the cluster center ($\sim0.1-0.5r_{\rm 200,cl}$) than early-type galaxies.
The AGN fraction for bright galaxies (C3) keeps increasing 
  even at $R<0.1r_{\rm 200,cl}$, 
  but there are only 2 and 20 galaxies 
  at $R\sim0.017r_{\rm 200,cl}$ and $0.056r_{\rm 200,cl}$,
  respectively, which shows a large uncertainty 
  (number of galaxies in other bins of C3 sample is $\sim90-2000$).
It is also seen that the AGN fraction for late types, 
  on average, is higher than that for early types in a given luminosity.
  
We wish to emphasize that it does not necessarily mean that
 the probability for cluster galaxies to be found as AGNs (i.e. $f_{\rm AGN}$) 
 is high when AGN host galaxies show an excess in the central region of clusters 
 as seen in top panels because the projected number density of non-AGN galaxies 
 also increases as decreasing clustercentric radius.

We now investigate the dependence of the AGN fraction
  on both clustercentric radius and nearest neighbor distance. 
We select a volume-limited sample of 
  target galaxies at $0.04\leq z<0.0927$
  whose absolute magnitudes are in a narrow range of $-20.0\geq M_r>-20.5$
  in order to reduce the effects of galaxy mass.
Dots in Fig. \ref{fig-fagn2d} show the distribution of target galaxies in the projected 
  clustercentric radius $R$ and projected nearest neighbor distance $R_n$ space.
Four panels distinguish among four different combinations of target and neighbor morphology. 
A spline kernel is used to obtain smooth distributions of the median $f_{\rm AGN}$ in 
  each location of the four panels. 
Contours with different colors mark constant AGN fractions.
If there are no strong effects of the distance to and the morphology of
  the neighbor galaxy on the AGN fraction,
  it is expected to see no significant difference in the shape of contours
  between left and right panels. However, we can see the difference.

Figure \ref{fig-fagn2d} shows that 
  the contours in the field (i.e. $R \ga 2 r_{\rm 200,cl}$) 
  are nearly horizontal in all panels.
This means that outside the cluster virial radius,
  the activity in galactic nuclei is determined strongly by 
  the nearest neighbor distance and morphology.
When the nearest neighbor of a galaxy is an early type (E-e and L-e case), 
  the AGN fraction decreases as the pair separation decreases at $R_n < r_{\rm vir,nei}$.
However, if the neighbor is a late type (E-l and L-l case), 
  it does not change much at $R_n > r_{\rm vir,nei}$
  but reaches a maximum at $R_n \sim 0.2 r_{\rm vir,nei}$.
This dependence of the activity in galactic nuclei on the neighbor galaxy in the field
  is consistent with the results in Choi et al. (in prep.).

\begin{figure*}
\center
\includegraphics[scale=0.63]{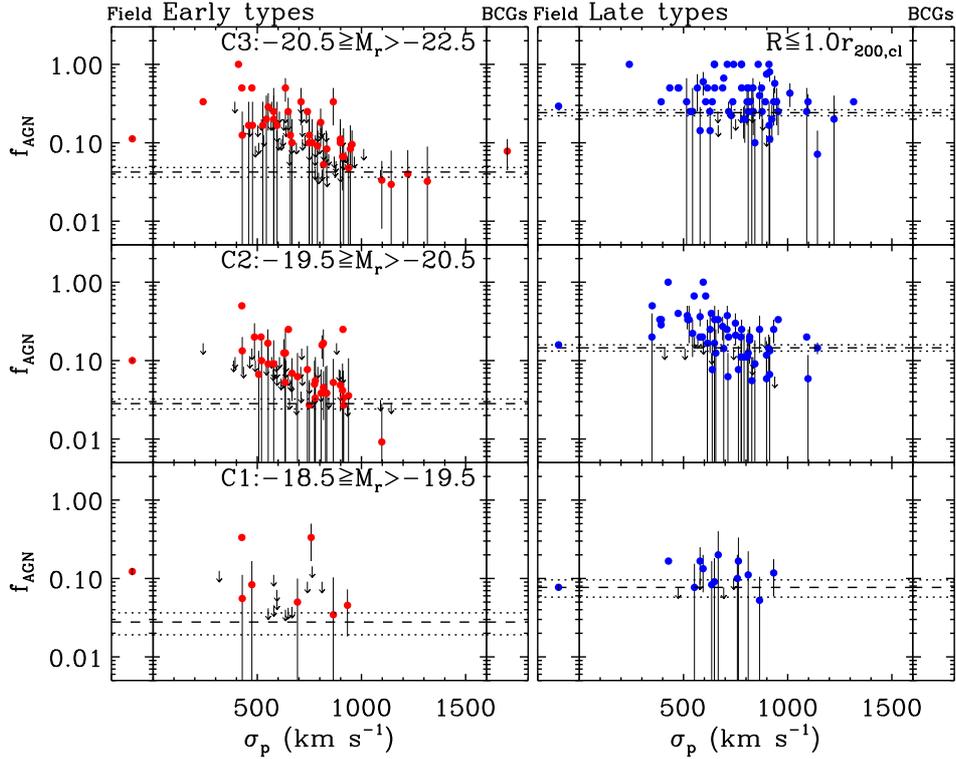}
\caption{AGN fraction ($f_{\rm AGN}$) as a function
  of velocity dispersion of galaxy clusters.
Left and right panels are for early- and late-type galaxies, respectively.
$f_{\rm AGN}$ is computed in the sample of galaxies 
  [C1({\it bottom})$-$C3({\it top})] within the
  virial radius of the clusters ($R\leq r_{\rm 200,cl}$).
Clusters with $f_{\rm AGN}=0$ are indicated by arrows
 at the positions of upper limits.
Horizontal dashed line indicates the mean $f_{\rm AGN}$
  using all the cluster galaxies in the sample, and dotted lines are its errors.  
$f_{\rm AGN}$ in the sample of field galaxies 
  ($R>r_{\rm 200,cl}$) and of BCGs  
  is shown to the left and right of each panel, respectively.
}\label{fig-fagnvsig}
\end{figure*}

\begin{table*}
\begin{center}
\caption{AGN fraction for subsamples and the correlation statistics\label{tab-sub}}
\begin{tabular}{cccccccccccc}
\hline\hline 
       & Early types   & &      &        &       && Late types    & &      &        & \\
\cline{2-6} \cline{8-12}
Sample & 
  $f_{\rm AGN,Field}$ & $f_{\rm AGN,Cluster}$ & $f_{\rm AGN,BCGs}$ & $\rho^{a}$ & Prob.$^{b}$ && 
  $f_{\rm AGN,Field}$ & $f_{\rm AGN,Cluster}$ & $f_{\rm AGN,BCGs}$ & $\rho^{a}$ & Prob.$^{b}$  \\
\hline
C3(Br) & 0.112$\pm$0.004 & 0.042$\pm$0.006 & 0.078$\pm$0.034 & $-0.81$ & 0.000 && 0.293$\pm$0.007 & 0.242$\pm$0.021 & ... & $-0.26$ & 0.038 \\
C2(Im) & 0.100$\pm$0.004 & 0.028$\pm$0.004 & ... & $-0.68$ & 0.000 && 0.159$\pm$0.004 & 0.146$\pm$0.013 & ... & $-0.63$ & 0.000 \\
C1(Fa) & 0.123$\pm$0.013 & 0.028$\pm$0.009 & ... & $-0.77$ & 0.043 && 0.077$\pm$0.007 & 0.077$\pm$0.019 & ... & $-0.34$ & 0.235 \\
\hline\hline
\end{tabular}
\begin{flushleft}
$^{a}$: The Spearman rank correlation coefficient.
$^{b}$: The two-sided probability of obtaining a value of $\rho$ by chance.
\end{flushleft}
\end{center}
\end{table*}

In the cluster region of $R< r_{\rm 200,cl}$,
  the shape of contours is relatively noisy, 
  but clearly different from that outside the cluster region.
Now all panels show somewhat slant contours.
Although early-type galaxies have late-type neighbors,
  $f_{\rm AGN}$ does not increase
  as they approach neighbor galaxies and the center of clusters.
When late-type galaxies have late-type neighbors,
  the increase of $f_{\rm AGN}$ is not significant 
  as they approach neighbor galaxies.
     
In Fig. \ref{fig-ed1d}, 
  we show $L_{\rm [OIII]}/M_{\rm BH}$ (AGN power indicator)
  as a function of the projected clustercentric radius $R$
  divided into two
  morphology and three magnitude bins.
It is seen that $L_{\rm [OIII]}/M_{\rm BH}$ for late types, on average, is larger
  than those for early types with similar luminosities.
The median values of  $L_{\rm [OIII]}/M_{\rm BH}$ do not
  show any noticeable change with $R$ for all subsamples.

In Fig. \ref{fig-edm}, we plot $L_{\rm [OIII]}/M_{\rm BH}$ again, 
  but as a function of stellar mass of host galaxies.
Now a strong anti-correlation between $L_{\rm [OIII]}/M_{\rm BH}$ and the stellar mass
  is seen for early-type AGNs,
  which means that massive AGNs currently
  accret less gas per unit black hole mass
  than less massive AGNs.
The Spearman rank correlation test for this correlation 
  gives a correlation coefficient of $-0.42$,
  and the two-sided probability of finding a value of $-0.42$ by chance is $<0.1$\%.
This anti-correlation was similarly seen in the plot of 
  $L_{\rm [OIII]}/M_{\rm BH}$ versus the black hole mass in previous studies \citep{heck04,best05}.
They suggested that  
  massive black holes ($M_{\rm BH}>$a few $10^7$ $M_\odot$) must have
  grown much faster in the past.
However, note that this trend might be because of ``Eddington incompleteness'', 
  which indicates an observational selection effect
  that AGNs with lower Eddington ratio are detected only at more massive galaxies
  in a given flux or luminosity limit (see \citealt{leegh11bar} for more discussion).
Note also that a weak correlation for late-type AGNs (right panel)
  may be due to a large uncertainty in $L_{\rm [OIII]}/M_{\rm BH}$.
The black hole mass is known to correlate with
  the velocity dispersion of bulge component, 
  but $M_{\rm BH}$ of late-type galaxies in this study
  was derived from the SDSS spectra obtained by a fixed size of optical fiber,
  which may introduce a systematic uncertainty in $M_{\rm BH}$.
In summary, the comparison between Figs. \ref{fig-ed1d} and \ref{fig-edm} 
  can tell us that the AGN power
  is strongly controlled by the mass of host galaxies (or SMBHs) rather than the environment.

\begin{figure*}
\center
\includegraphics[scale=0.61]{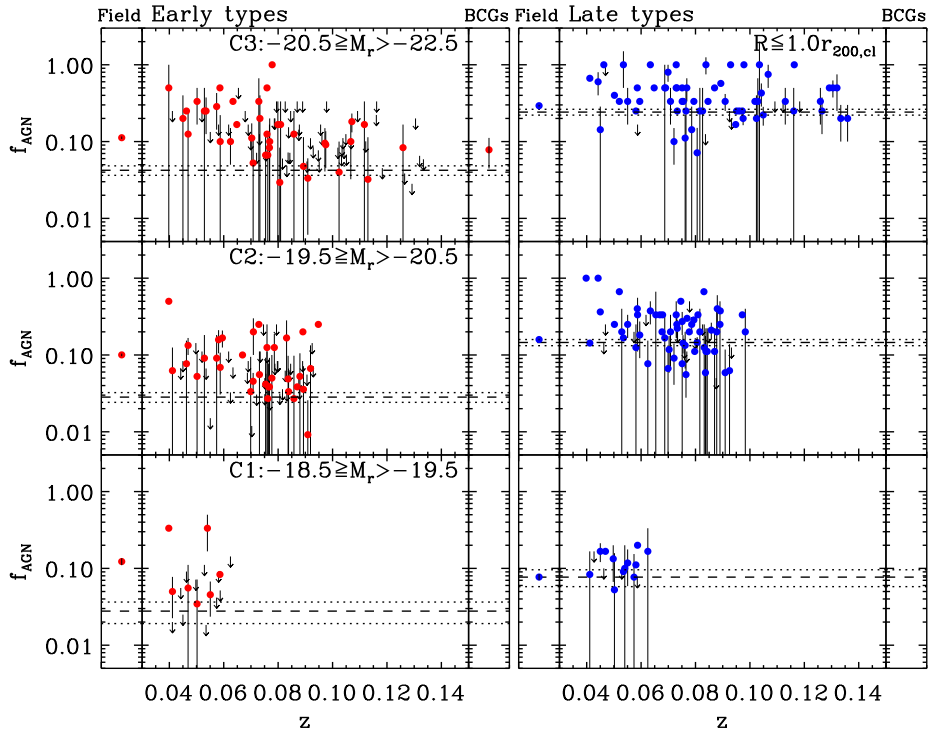}
\caption{Same as Fig. \ref{fig-fagnvsig},
  but for $f_{\rm AGN}$ as a function
  of redshift of galaxy clusters.
}\label{fig-fagnz}
\end{figure*}

\subsection{AGN fraction depending on clusters}

In Figs. \ref{fig-fagnvsig} and \ref{fig-fagnz},
  we show the AGN fraction of each cluster
  as a function of cluster velocity dispersion and redshift, respectively.
The AGN fraction of each cluster is computed using the galaxies within 
  one virial radius of the cluster ($R\leq r_{\rm 200,cl}$),
  and its error bar represents 68\% (1$\sigma$) confidence interval
  that is determined by the bootstrap resampling method.
It is seen that the AGN fraction decreases 
  as the velocity dispersion of cluster
  increases when we consider only the clusters with $f_{\rm AGN}>0$.
It is prominent for the sample of C2 and C3.
To check this correlation between the two quantities 
  [log($f_{\rm AGN}$) vs. $\sigma_{\rm p}$],
  we compute the Spearman rank correlation coefficient ($\rho$) and its significance 
  using the clusters with $f_{\rm AGN}>0$ in each panel, 
  and summarize the results in Table \ref{tab-sub}.
It shows that the anti-correlation between the two
  really exist for all subsamples 
  divided by luminosity and morphology of host galaxies,
  and the anti-correlation is stronger in early types than in late types.
This finding is consistent with the results in previous studies 
   (e.g., \citealt{pb06,arn09}),
   but we further show that this anti-correlation exists
   for all subsamples 
   divided by luminosity and morphology of host galaxies.

One might think that the AGN fractions in this figure 
  are somewhat large (sometimes reaches unity) 
  compared to the results in previous studies.
However, we would like to emphasize that
  the AGN fraction strongly depends on 
  the luminosity and the morphology of host galaxies,
  so simple comparison of AGN fractions
  without controlling physical parameters could be misleading.
Note again that the AGN fraction in this study is a ratio of 
  the number of Type I plus Type II AGNs 
  (Seyferts, LINERs and composite galaxies determined in the line ratio diagram)
  to the number of galaxies with spectroscopic parameters measured in MPA/JHU DR7 VAGC.
Therefore, the AGN fraction can be different
  if the galaxies with spectroscopic parameters 
  measured in MPA/JHU DR7 VAGC are not complete.
Indeed, the spectroscopic completeness of SDSS data
  is as low as 80\% in high-density regions 
  such as central regions of clusters ($\sim$90\% in the field)
  because of difficulty in observing with multi-object spectrograph
  (mostly early-type galaxies or non-emission line galaxies; 
  see \S\ref{sdss} and Fig. 1 in \citealt{ph09}),
  which can affect our results slightly.
   
Comparison of the mean AGN fraction computed using all the cluster galaxies 
  ($R\leq r_{\rm 200,cl}$)
  with that of field galaxies shows that
  the AGN fraction of early-type cluster galaxies is lower than 
  that of early-type field galaxies
  by a factor $\sim 3$ for all luminosity subsamples (see Table \ref{tab-sub}).
The AGN fraction of field galaxies is computed using the galaxies
  outside the cluster region ($R>r_{\rm 200,cl}$).  
However, the AGN fractions of late-type cluster galaxies
  are, on average, not different from those of late-type field galaxies.
The AGN fraction of early-type BCGs in C3 sample (5 out of 64) 
  appears to be marginally smaller than that of early-type field galaxies
  with similar luminosities (see top left panel in Fig. \ref{fig-fagnvsig}),
  but be slightly higher than that of other early-type cluster galaxies 
  with similar luminosities.
   
When we plot the variation of AGN fraction as a function of cluster redshift 
  (Fig. \ref{fig-fagnz}),
  we do not see any significant dependence of $f_{\rm AGN}$ on redshift.
However, note that the redshift range in this study is much narrower than
  other studies that show a substantial increase of AGN fraction of cluster galaxies
  (e.g., $0.0\la z \la 1.3$ in \citealt{mar09} and \citealt{tom11}).

\subsection{Comparison of physical parameters between AGN hosts and non-AGN galaxies}

\begin{figure}
\center
\includegraphics[scale=0.35]{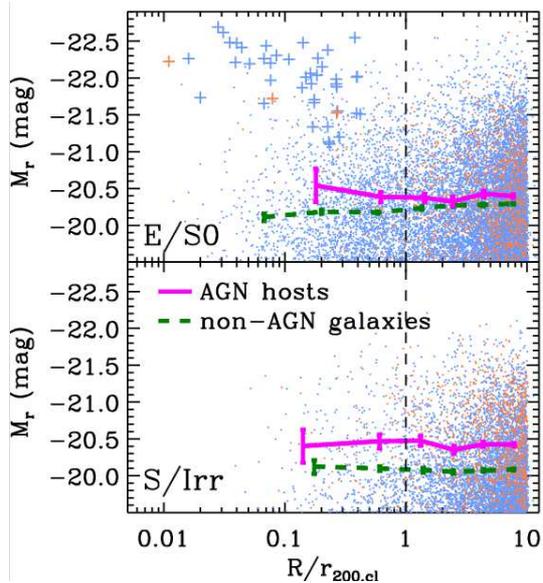}
\caption{Absolute magnitude of galaxies brighter than 
  $M_r=-19.5$ with $0.04\leq z < 0.0927$
 vs. clustercentric radius. 
The upper panel shows early types, 
  and the lower panel shows late types. 
Orange and blue dots indicate AGN hosts and non-AGN galaxies, respectively.
Solid lines are median magnitudes for AGN host galaxies,
  while dashed lines are those for non-AGN galaxies.
Crosses are BCGs, and are not used in calculating median curves.
Late types with axis ratio of $b/a<0.6$ are eliminated.
Vertical dashed lines indicate the cluster virial radius $r_{\rm 200,cl}$.
}\label{fig-mag1d}
\end{figure}

Figure \ref{fig-mag1d} shows $r$-band absolute magnitudes of 
  AGN hosts and non-AGN galaxies in a 
  volume-limited sample of $0.04\leq z < 0.0927$ and $M_r < -19.5$. 
The lines with error bars are median values as a function of $R$. 
The BCGs, marked as crosses, are not used in calculating median values.
It is seen that absolute magnitudes for both AGN hosts and non-AGN galaxies
  do not change much with the clustercentric radius.
We call non-AGN galaxies those objects 
  whose optical spectral types are 
  star-forming, ambiguous galaxies, or undetermined.
Interestingly,
  AGN host galaxies are systematically brighter than non-AGN galaxies
  by $\sim0.2$ and $0.4$ mag for early and late types, respectively.

\begin{figure*}
\center
\includegraphics[scale=0.65]{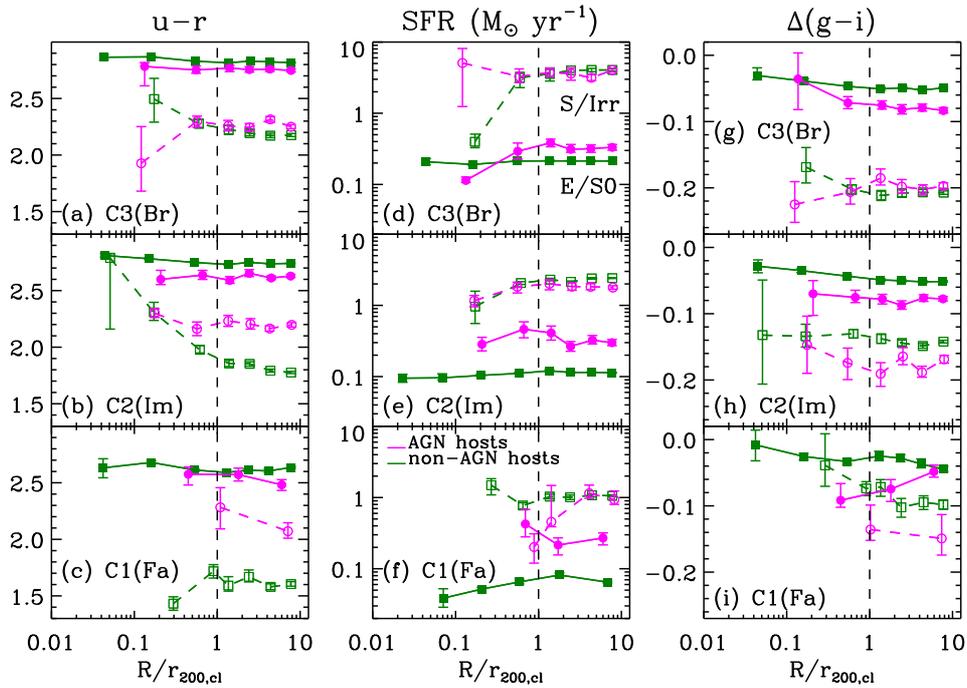}
\caption{Dependence of physical parameters of our target galaxies 
  in the samples of C1$-$C3 on the clustercentric radius:
  ({\it left}) ($u-r$), ({\it middle}) SFR, and 
  ({\it right}) $\Delta(g-i)$.
Median curves are drawn for the cases of 
  early-type AGN host galaxies (filled circle and solid line),
  early-type non-AGN galaxies (filled square and solid line),
  late-type AGN host galaxies (open circle and dotted line), and
  late-type non-AGN galaxies (open square and dotted line).
Late types with axis ratio of $b/a<0.6$ are eliminated.
SFRs are computed with $H_0 = 70$ km s$^{-1}$ Mpc$^{-1}$ and Salpeter IMF.
}\label{fig-sf1d}
\end{figure*}

In Fig. \ref{fig-sf1d}, 
  we plot ($u-r$) color, SFR, and ($g-i$) color gradient
  of AGN hosts and non-AGN galaxies divided into two
  morphology and three magnitude bins as a function of $R$.
It is seen that ($u-r$) colors for AGN host galaxies do not change much
  as $R$ decreases regardless of the luminosity and morphology
  except for C2 late-type hosts.
Comparison between AGN hosts and non-AGN galaxies shows that 
  early-type AGN hosts tend to be bluer than early-type non-AGN galaxies 
  at all clustercentric radii,
  while late-type AGN hosts are redder than late-type non-AGN galaxies,
  which is consistent with previous results 
  (e.g., \citealt{kau03agn,choi09,sch10host}).
However, the difference between AGN hosts and non-AGN galaxies
  is not significant for the brightest sample (C3).

The SFRs of late-type AGN hosts
  are not distinguishable from those of late-type non-AGN galaxies
  except for the brightest sample (C3) 
  at the very inner center ($R\lesssim0.2r_{\rm 200,cl}$).
For early types,
  SFRs of AGN host galaxies seem to be larger than those of non-AGN galaxies 
  at all clustercentric radii,
  but the difference for the brightest sample (C3) is small.

\begin{figure*}
\center
\includegraphics[scale=0.65]{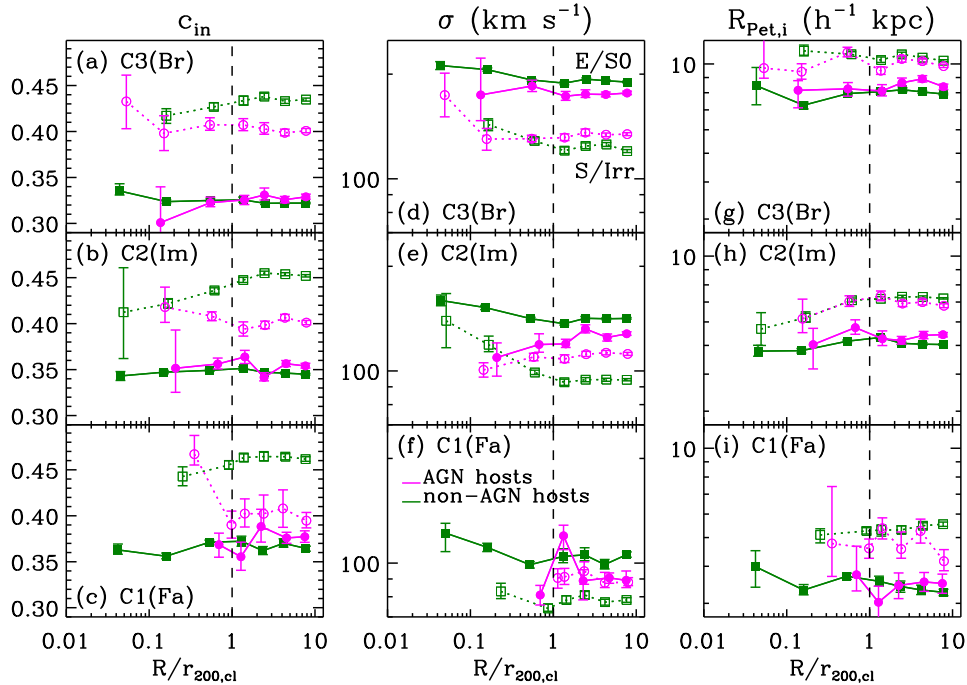}
\caption{Same as Fig. \ref{fig-sf1d}, but for
  ({\it left}) $c_{\rm in}$, ({\it middle}) $\sigma$ (km s$^{-1}$), and 
  ({\it right}) R$_{\rm Pet,i}$ ($h^{-1}$ kpc).
Late-types with axis ratio of $b/a<0.6$ are included.
}\label{fig-st1d}
\end{figure*}

($g-i$) color gradients for non-AGN galaxies increases (redder core) as $R$ decreases,
  while the dependence of those for AGN host galaxies on $R$ is very weak.
Comparison between AGN hosts and non-AGN galaxies shows that
  the color gradients for early-type AGN hosts are smaller than 
  those for early-type non-AGN galaxies, which might imply that
  the nuclear activity makes the galaxy core bluer.

Figure \ref{fig-st1d} represents the structure parameters such as 
  concentration index ($c_{\rm in}$),
  internal velocity dispersion ($\sigma$), and
  Petrosian radius in $i$-band of AGN hosts and non-AGN galaxies divided into two
  morphology and three magnitude bins as a function of $R$.

The left column of Fig. \ref{fig-st1d} does not show any noticeable change
  in the concentration index of AGN host galaxies with the clustercentric radius.
Comparison of AGN hosts with non-AGN galaxies shows that
  late-type AGN hosts are more likely centrally 
  concentrated (i.e. $c_{\rm in}$ is smaller) than 
  late-type non-AGN galaxies. 
However, any significant difference is not seen for early-type galaxies.

It is seen that internal velocity dispersions of non-AGN galaxies
  increase as decreasing $R$ within the cluster region, 
  but AGN hosts do not show a similar behavior.
The velocity dispersions for early-type AGN hosts, on average, 
  appear to be marginally smaller than those for early-type non-AGN galaxies, 
  but those for late-type AGN hosts
  tend to be larger than those for late-type non-AGN galaxies.
  
In the right column of Fig. \ref{fig-st1d}, we do not see any noticeable
  change of Petrosian radius of AGN host galaxies with $R$
  and any significant difference between AGN hosts and non-AGN galaxies.

In summary, 
  the difference in physical parameters
  such as ($u-r$) colors, SFRs, and ($g-i$) color gradients
  between AGN hosts and non-AGN galaxies
  seems to exist for both early and late types at all clustercentric radii, while 
  the difference in structure parameters between the two
  is significant only for late types.
However, the dependence of these parameters for AGN host galaxies
  on the clustercentric radius is not clearly distinguishable from 
  that for non-AGN galaxies.
    
In particular, the difference in concentration index 
  and internal velocity dispersion 
  between late-type, AGN hosts and non-AGN galaxies (Fig. \ref{fig-st1d}),
  are significant at all clustercentric radii, 
  while early types do not show any noticeable difference
  (see also \citealt{choi09}).
This seems to be consistent with the idea that
  massive bulges that are highly concentrated compared to disks 
  and have high velocity dispersions,
  are necessary for having SMBHs and their activity \citep{fm00,geb00}.

\section{Discussion}\label{discuss}

\subsection{Activity in galactic nuclei in galaxy clusters and in the field}\label{discussenv}

From a point of view of fuel supply to AGNs,
  it is important to know the gas content of cluster and field
  galaxies in order to understand the difference between the two.
For example, it was known that the molecular gas content of cluster, late-type galaxies
  is not different from that of field, late-type galaxies 
  (\citealt{bg06} and references therein).
  This is similarly found for early-type galaxies \citep{young11}.
However, the atomic gas content of cluster galaxies
  is found to be lower than that of field galaxies (e.g. \citealt{dav73,di07}).
This is generally interpreted as that
  the molecular gas that is denser and more deeply embedded in a galaxy potential well
  than the atomic gas,
  is not easily removed by any cluster-related 
  stripping mechanism (see \citealt{bg06} for a review). 

In the previous section, 
  we found that the AGN fraction of early-type galaxies starts to decrease
  around $r_{\rm 200,cl}$ as decreasing clustercentric radius,
  and AGN fractions of early-type cluster galaxies ($R\leq r_{\rm 200,cl}$) 
  are, on average, lower than
  those of field galaxies by a factor of $\sim 3$ 
  (see Figs. \ref{fig-fagn1d} and \ref{fig-fagnvsig}), 
  which is consistent with previous studies (e.g. \citealt{arn09,von10}).
On the other hand,
  the AGN fractions of late types decrease
  much closer to the cluster center ($\sim0.1-0.5r_{\rm 200,cl}$) than those of early types,
  and the AGN fractions of late-type cluster galaxies are, 
  on average, not different 
  from mean AGN fractions of field late types.

Firstly, the gradual decline of the AGN fraction in both types
  means that the effects of physical mechanisms on triggering (quenching) 
  the activity in galactic nuclei
   are inefficient (stronger) in the central region of clusters.  
Secondly, the different behavior of the AGN fraction 
  depending on the morphology of host galaxy,
  indicates an important role of host morphology in the nuclear activity,
  especially for cluster galaxies.
Cluster galaxies make orbital motions in a cluster potential,
  but their orbits are known to be different depending
  on galaxy morphology \citep{bk04,hl08}.  
For example,
  early-type cluster galaxies are found to be in isotropic orbits,
  being in equilibrium in the cluster potential 
  through several orbiting motions in the past,
  which results in no cold gas (atomic and molecular gas) left in them.
On the other hand, late-type galaxies are in radial orbits,
  which suggests that they enter the cluster region recently.
Therefore, their cold gas (mainly molecular gas) may not be stripped or consumed yet, and
  they can still feed their SMBHs as in the field \citep{mar09,hag10}.

Interestingly, 
  we found the anti-correlation between $f_{\rm AGN}$ and $\sigma_{\rm p}$ in clusters
  as seen in previous studies (e.g., \citealt{pb06,arn09}), 
  but for all subsamples divided by morphology and luminosity of host galaxies.
Since we fix the host morphology and luminosity,
  this anti-correlation is not simply due to the different morphological mix 
  depending on the environment.
This result can be explained if AGNs are triggered through galaxy-galaxy interactions
  and mergers because the probability for galaxy-galaxy interactions and mergers
  is expected to be low when the relative velocity between galaxies is high
  in massive clusters having large velocity dispersions 
  (e.g., \citealt{mamon92,ghi98,hop08I,hop08II}).
It is also consistent with the result that galaxies in more X-ray luminous clusters
  have less cold gas that is necessary for feeding SMBHs (e.g., \citealt{gh85}).
  
This interaction induced nuclear activity
  can also explain the decrease of the AGN fraction as decreasing clustercentric radius
  seen in Fig. \ref{fig-fagn1d} as follows.
As galaxies with SMBHs approach the cluster center,
  the probability for galaxies to interact with other ``late-type'' galaxies
  decreases, which is necessary for making the SMBHs active.
Similarly, as galaxies approach the cluster center,
  they experience continuous interactions with
  other galaxies (mostly early types) 
  because the virial radii of galaxies all overlap with one another
  within the viral radius of cluster (see Fig. \ref{fig-fagn2d}).
Then the nearest neighbors (mostly early types) may have left significant cumulative effects
  of decreasing activity in galactic nuclei (to be discussed in \S \ref{power}). 
The amount of the cumulative effects increases monotonically as the clustercentric
  radius decreases, which can produce the trend seen in Fig. \ref{fig-fagn1d}.

Interestingly, 
  the AGN fraction of BCGs is found to be marginally 
  higher than that of other cluster galaxies 
  with similar luminosities (see top left panel in Fig. \ref{fig-fagnvsig})
  even if they are all early types,
  but be slightly lower than that of field early-type galaxies.
Previously, it was suggested based on SDSS C4 cluster catalog \citep{mil05}
  that BCGs are more likely to host radio-loud AGNs, but less likely
  to host (powerful) optical AGNs than other galaxies with similar masses
  (e.g., \citealt{best07,von07}).
The different results between this and their studies
  might be due to
  a different mass range of cluster sample 
    (our Abell clusters are, on average, more massive than C4 clusters,
    and in fact three out of five BCGs with the activity in galactic nuclei in C3 sample
    are found in clusters with $\sigma_p>660$ km s$^{-1}$),
  a different morphological galaxy sample 
    (galaxies are compared by fixing morphology in this study), and/or
  a different comparison sample of BCGs 
    (we divided the comparison sample into two: field and other cluster galaxies).
On the other hand,
  the result in this study is consistent with that in previous studies
  in the sense of enhanced nuclear activity of BCGs compared to other cluster galaxies,
  which may be directly related to 
  the cooling of X-ray gas at the cluster center 
  (e.g., \citealt{burns90,best07,edw07,san09}).
However, among five clusters with the activity in galactic nuclei of BCGs in this study
  (A795, A1171, A1213, A1668, and A1991),
  there is clear evidence of cooling flow for only one cluster (A1991; \citealt{sha04}).

\subsection{What triggers activity in galactic nuclei?}\label{power}
  
When we consider the dependence of AGN fraction
  on the nearest neighbor galaxies as well as clusters,
   the activity in galactic nuclei outside the cluster virial radius (i.e. in the field)
   is found to be determined strongly by 
   the morphology of and distance to the nearest neighbor 
   when morphology and luminosity of host galaxy are fixed
   (see Fig. \ref{fig-fagn2d}; Choi et al. in prep.).
When the nearest neighbor of a galaxy is an early type (E-e and L-e case), 
  the AGN fraction decreases as the pair separation decreases 
  at $R_n < r_{\rm vir,nei}$.
However, if the neighbor is a late type (E-l and L-l case), 
  it does not change much at $R_n > r_{\rm vir,nei}$
  but reaches a maximum at $R_n \sim 0.2 r_{\rm vir,nei}$.
This dependence of $f_{\rm AGN}$
  on the morphology of and the distance to the nearest neighbor galaxy 
  in the field, can support the idea that AGNs are triggered
  by galaxy-galaxy interactions and mergers.

In particular, the different behavior of $f_{\rm AGN}$
  depending on the morphology of the nearest neighbor
  starts to be seen at $R_{\rm n}\approx r_{\rm vir,nei}$
  where the galaxies in pair start to interact hydrodynamically \citep{park08}.
This can imply that hydrodynamic interactions
  with the nearest neighbor play important roles in triggering nuclear activity
  in addition to the tidal interactions.
If a galaxy with a SMBH approaches a late-type neighbor 
  within the virial radius of the neighbor,
  the inflow of cold gas from the neighbor into the target galaxy increases and
  the SMBH starts to accrete the gas and to be active.
The crossing time of galaxies across the virial radius is
  of an order of $\sim10^9$ yrs, 
  which is much shorter than the age of the universe.
Therefore, this inflow is expected to occur for $\lesssim10^9$ yrs, 
  which is also seen in the simulations (e.g., \citealt{hop08I}). 
The mass transfer between galaxies in pair is usually observed
  in close pairs with a pair separation of $\lesssim$30 kpc 
  (e.g., \citealt{kew10, font11}).
There are some candidates found in the SDSS images
  with large pair separations (see Fig. 4 in \citealt{park08}).
Moreover, there is a known ultraluminous infrared galaxy
  with a pair separation of $\sim$90kpc,
  which shows nuclear activity and large tidal features 
  ({\it IRAS} 11223-1244; \citealt{dckim02}),
  which also supports our argument.
    
On the other hand, if an early-type galaxy approaches an early-type neighbor 
  within the virial radius of the neighbor even if it has a SMBH,
  the engine (SMBH) would not be ignited due to the lack of fuel (gas supply),
  which results in the low value of $f_{\rm AGN}$.
When a late-type galaxy with a SMBH approaches an early-type neighbor 
  within the virial radius of the neighbor,
  the activity in galactic nuclei can be sustained since
  the gravitational interaction with the neighbor makes the gas in the host galaxy
  move toward the center to feed SMBH even if there is no gas inflow from the neighbor. 
Therefore, if we do not consider the morphology of target and neighbor galaxies,
  the different role of galaxy morphology will be averaged out.
This can explain why the dependence of AGN fraction on the distance to 
  the neighbor galaxies is not seen in some previous studies.

Within the cluster region of $R\lesssim r_{\rm 200,cl}$,
  galaxies start to be affected by the cluster itself and its member galaxies.
The cluster galaxies are expected to 
   experience repeated gravitational or hydrodynamical interactions with other galaxies 
   (and with cluster itself)
   as they make trapped orbital motions within the cluster 
   (see \citealt{ph09,bg06} for a review).
However, the orbital velocity of cluster galaxies are 
  very high and the tidal energy deposit during the short encounters 
  is too small to significantly affect galaxy properties \citep{mer84,bv90,ph09}.
Therefore, though cluster galaxies are found to have current close neighbor galaxies,
  galaxy properties that we observe now may reflect 
  the cumulative gravitational or hydrodynamical effects 
  that they have experienced during the orbital motion.
This can weaken the dependence of galaxy properties on 
  the nearest neighbor galaxies in the cluster region,
  and may explain why $f_{\rm AGN}$ of cluster galaxies (at $R<r_{\rm 200,cl}$) 
  does not increase unlike field galaxies
  when they approach 
  ``late-type'' neighbors at  $R_{\rm n}<r_{\rm vir,nei}$
  as seen in right panels of Fig. \ref{fig-fagn2d}.
They might have experienced several gravitational or hydrodynamical interactions
  with cluster galaxies that are mainly early types, which results in the lack of gas
  to feed SMBHs.

\section{Conclusions}\label{con}

Using the SDSS data,
  we have studied the environmental dependence of the activity in galactic nuclei
  by comparing cluster and field galaxies,
  and have compared the galaxy properties of AGN hosts and non-AGN galaxies.
Our main results follow.

\begin{enumerate}
\item The AGN fraction of early-type galaxies starts to decrease
  around one virial radius of clusters as decreasing clustercentric radius,
  while that of late types starts to decrease
  close to the cluster center ($R\sim0.1-0.5 r_{\rm 200,cl}$).

\item When we consider the dependence of AGN fraction 
  not only on the clustercentric radius
  but also on the morphology of and the distance to the nearest neighbor,
  it appears to depend on both quantities in cluster regions.
On the other hand, in the field,
  the AGN fraction is found to  
  depend strongly on the morphology of and 
  the distance to the nearest neighbor galaxy 
  when morphology and luminosity of host galaxy are fixed.

\item The AGN fractions of early-type cluster galaxies, on average,
  are lower than those of early-type field galaxies
  by a factor $\sim 3$ for all luminosity subsamples.
However, mean AGN fractions of late-type cluster galaxies
  are not different from those of late-type field galaxies.
The different accretion history between early and late types
  might explain this difference.

\item We found the anti-correlation 
  between $f_{\rm AGN}$ and $\sigma_{\rm p}$ in clusters
  for all subsamples divided by morphology and luminosity of host galaxies.
 
\item The difference in physical parameters
  such as luminosity, ($u-r$) colors, SFRs, and ($g-i$) color gradients
  between AGN hosts and non-AGN galaxies
  is seen for both early and late types at all clustercentric radii, while 
  the difference in structure parameters between the two
  is significant only for late types.
However, the changes of these parameters of AGN host galaxies
  with the clustercentric radius are not clearly distinguishable from those of non-AGN galaxies.

\item $L_{\rm [OIII]}/M_{\rm BH}$ (AGN power indicator)
  for late types, on average, is larger
  than those for early types with similar luminosities.
It is also found that the AGN power
  is weakly dependent on the clustercentric radius,
  but strongly dependent on the mass of host galaxies.
   
\item The AGN fraction of early-type BCGs is found to be 
  marginally higher than
  that of other early-type cluster galaxies with similar luminosities,
  but be slightly smaller than that of early-type field galaxies.
  The enhanced nuclear activity of BCGs compared to other cluster galaxies
  may be related to the cooling of X-ray gas at the cluster center. 
  

\end{enumerate}

Our results are consistent with the idea
  that the activity in galactic nuclei is triggered through galaxy-galaxy interactions and mergers
  when the fuel (gas supply) for AGN is available.
   
\begin{acknowledgements}
We would like to thank the anonymous referee 
  for constructive comments that helped improving the paper,
  and R. Gobat, J. Mullaney, and G.-H. Lee for useful discussion.
H.S.H. acknowledges the Centre National d`Etudes Spatiales (CNES)
  and the Smithsonian Institution for the support of his post-doctoral fellowship.
D.E. acknowledges the support of the Centre National d`Etudes Spatiales (CNES).
Funding for the SDSS and SDSS-II has been provided by the Alfred P. Sloan 
Foundation, the Participating Institutions, the National Science 
Foundation, the U.S. Department of Energy, the National Aeronautics and 
Space Administration, the Japanese Monbukagakusho, the Max Planck 
Society, and the Higher Education Funding Council for England. 
The SDSS Web Site is http://www.sdss.org/.
The SDSS is managed by the Astrophysical Research Consortium for the 
Participating Institutions. The Participating Institutions are the 
American Museum of Natural History, Astrophysical Institute Potsdam, 
University of Basel, Cambridge University, Case Western Reserve University, 
University of Chicago, Drexel University, Fermilab, the Institute for 
Advanced Study, the Japan Participation Group, Johns Hopkins University, 
the Joint Institute for Nuclear Astrophysics, the Kavli Institute for 
Particle Astrophysics and Cosmology, the Korean Scientist Group, the 
Chinese Academy of Sciences (LAMOST), Los Alamos National Laboratory, 
the Max-Planck-Institute for Astronomy (MPIA), the Max-Planck-Institute 
for Astrophysics (MPA), New Mexico State University, Ohio State University, 
University of Pittsburgh, University of Portsmouth, Princeton University,
the United States Naval Observatory, and the University of Washington. 
This research has made use of the NASA/IPAC Extragalactic Database (NED) 
which is operated by the Jet Propulsion Laboratory, California Institute of Technology, 
under contract with the National Aeronautics and Space Administration.
\end{acknowledgements}

\bibliographystyle{aa} 
\bibliography{ref_hshwang} 
\end{document}